\documentclass[MSNbibl,nameyear,dvips]{arxstspdf}
\usepackage{flushend}
\usepackage{stfloats}
\usepackage{graphicx}

\volume{26}
\issue{4}
\pubyear{2011}
\firstpage{564}
\lastpage{583}
\doi{10.1214/11-STS375}

\makeatletter
\fnbelowfloat
\newcommand{\EE}{\mathbb{E}}

\makeatother

\begin{document}
\begin{frontmatter}

\title{Forensic Analysis of the Venezuelan Recall Referendum}
\runtitle{Forensics of the Venezuelan Referendum}

\begin{aug}
\author{\fnms{Ra\'ul} \snm{Jim\'enez}\corref{}\ead[label=e1]{rauljose.jimenez@uc3m.es}}
\runauthor{R. Jim\'enez}

\affiliation{Universidad Carlos III de Madrid}

\address{Ra\'ul Jim\'enez is Associate Professor, Department of Statistics,
Universidad Carlos III de Madrid,
C./ Madrid, 126 -- 28903 Getafe (Madrid), Spain
\printead{e1}.}

\end{aug}

\begin{abstract}
The best way to reconcile political actors in a controversial electoral process is a full audit.
When this is not possible, statistical tools may be useful for measuring the likelihood of the results.
The Venezuelan recall referendum (2004) provides a suitable dataset for thinking about this important problem.
The cost of errors in examining an allegation of electoral fraud can be enormous.
They can range from legitimizing an unfair election to supporting an unfounded accusation, with serious political implications.
For this reason, we must be very selective about data, hypotheses and test statistics that will be used.
This article offers a critical review of recent statistical literature on the Venezuelan referendum.
In addition, we propose a testing methodology, based exclusively on vote counting,
that is potentially useful in election forensics.
The referendum is reexamined, offering new and intriguing aspects to previous analyses.
The main conclusion is that there were a~significant number of irregularities in the vote counting
that introduced a bias in favor of the winning option.
A plausible scenario in which the irregularities could overturn the results is also discussed.
\end{abstract}


\end{frontmatter}

\section{Introduction}\vspace*{3pt}

The statistical controversies surrounding the outcomes of the
Venezuelan referendum, convened to revoke the mandate of President Ch\'
avez on August 15th of 2004, generated a long spate of articles in
newspapers and occupied significant television time.
A Google search with the exact phrase ``Venezuelan recall referendum''
shows more than 100,000 hits in English.
Several reports, commissioned by different organizations, reached
opposite conclusions.
Roughly speaking, a fraud\vadjust{\goodbreak} may have occurred during the referendum or,
on the contrary, was statistically undetectable.
A~good example of this is the work of Hausmann and Rigobon (\citeyear{hausmann2011}), where
the authors claimed to have found statistical evidence of fraud.
According to experts consulted by \textit{The Wall Street Journal}, ``the
Hausmann/Rigobon study is more credible than many of the other
allegations being thrown around'' (Luhnow and De Cordoba, \citeyear{luhnow2004}).
However, their early claim (Hausmann and Rigobon, \citeyear{hausmann2004})
was later rejected by The Carter Center [(\citeyear{cc2005}), Appendix 4] and
by Weisbrot et al. (\citeyear{cepr2004}).

The first peer-reviewed article devoted to the statistical analysis of
the referendum data (Febres and Marquez, \citeyear{febres2006})
concluded that there is statistical evidence for rejecting the official results.
This article, in \textit{International Statistical Review}, made no
mention of the paper by Taylor (\citeyear{taylor})
which concluded explicitly that there is no evidence of fraud.
Taylor's paper is the best known reference on the subject, widely
covered by media; in part because he was\vadjust{\goodbreak} asked to investigate the
allegations of fraud on behalf of The Carter Center.
Another well-known reference is a paper by Felten et al. (\citeyear{felten2004}), which
did not detect any statistical inconsistency that would indicate
obvious fraud in the election.
However, three papers
in this issue of \textit{Statistical Science} (Delfino and Salas,
\citeyear{delfino}; Prado and Sans\'o, \citeyear{prado}; Pericchi
and Torres, \citeyear{pericchi}) support the claim of fraud.
Who is right?

The statistical papers on the referendum can be grouped into two classes:
those that only use vote counting and those that use related additional data.
Five papers mentioned above cover the different\break claims of fraud
investigated by the panel of experts convened by The Carter Center
[(\citeyear{cc2005}), Appendix~13].
These are:
\begin{longlist}
\item[(1)] Discrepancy between official results and exit polls (Prado and
Sans\'o, \citeyear{prado}) and unexpected correlations between
computerized vote counting, the number of signatures for the recall
petition and audit results (Delfino and Salas, \citeyear{delfino}).
\item[(2)] Anomalous distributions of votes among voting notebooks
(Febres and Marquez \citeyear{febres2006}; Taylor \citeyear{taylor}), including high rates of ties (Taylor, \citeyear
{taylor}) and failure of fit to \textit{Benford's Law} for significant
digits (Pericchi and Torres \citeyear{pericchi}; Taylor \citeyear{taylor}).
\end{longlist}
I am very skeptical about the use of data from other sources.
To make a long story short, below I mention only key facts that can be
extracted from the Comprehensive Report of The Carter Center:

\textit{The months previous to the referendum were highly polarized, with
mass rallies for and against the government, with aggressive campaigns
for attracting new voters and to intimidate and persecute both signers
(people who signed for the recall petition) and supporters of President
Ch\'avez.
Even the referendum day was hot. The electoral actors took ad hoc
decisions that generated suspicions and lack of confidence in the whole
process.}

In this political atmosphere, we must assume that any unofficial
information will be controversial.
If there are many doubts about the official results, one cannot expect
consensus with other data.
Furthermore, one must be very careful with the statistical assumptions
that one will use.

This article has two purposes: (1) to bring order to the ruckus caused
by different statistical analyses, some of them carried out by
non-experts, and
(2) to examine, by a proper forensics analysis, the allegations of fraud.
Section \ref{sec2} reviews the referendum framework, introduces the main
notation used throughout this paper and presents a critical revision of
the five papers cited above.
In Section \ref{sec3} we propose a methodology, based exclusively on vote
counting, to test the recall referendum of 2004.
The presidential elections of 1998 and 2000 are also reviewed.
Far from being a statistical headache, the referendum is an excellent
dataset to exercise a wide variety of elementary but powerful
statistical tools.
Additionally, the case of study is also useful for illustrating some
common mistakes in stochastic modeling.
Section \ref{sec4} summarizes 
the main findings and conclusions.

\section{Referendum Framework and~Critical~Review}\label{sec2}

The electoral process is fully described in the report of The Carter
Center (\citeyear{cc2005}).
The crucial features for the present analysis are:

\begin{longlist}
\item[(i)]
\textit{A voting center consists of one or more electoral tables and each
table consists of one, two or three voting notebooks, which are the
official data units with the lowest number of votes.}
\item[(ii)]
\textit{Within the time allowed, voters were registered to a center.
Voters usually chose a center close to their residence or workplace,
many of them long before the referendum.
When the time was over, the referee decided the number of voting
notebooks in a center according to the number of voters.
In addition, notebooks are grouped in tables (no more than three per
table), mainly for logistical reasons related to the voting process.}
\item[(iii)]
\textit{In each center, voters are randomly assigned to the
notebooks}.\footnote{Every Venezuelan citizen is assigned an ID number.
These numbers are assigned in sequential order by date of request.
Usually, it is done when a Venezuelan girl or boy is ten years old.
By this I mean that the number is independent of the entire electoral process.
The ID number of the voters (older than 18 years old) has up to nine
digits and, except for a~case of extreme longevity, at least six digits.
The mechanism to assign voters to notebooks can be described as follows:
According to the last two digits, the voters were uniformly distributed
to the notebooks.
For example, in a~center with four notebooks,
if the last two digits ended between 00 and 24, then it was assigned to
notebook 1.
If the last two digits ended between 25 and 49, then it was assigned to
notebook 2, and so on.}
\item[(iv)]
\textit{There were only two options to vote: YES or NO.
Although there was a very small percentage of invalid votes (0.3\%),
there was a significant percentage of abstentions (30\%).}
\item[(v)]
\textit{The voting notebooks were computerized\break (touch-screen voting
machines which collected 86\% of the valid votes) and manual (ballot
boxes which represented 14\% of the votes).}
\end{longlist}

Both (i)--(ii) and (iv)--(v) are simple true facts but~(iii) is a \textit
{statistical hypothesis}.
The secrecy of the ballot lies in the random assignment of voters to notebooks.
For this reason, (iii) is essential for a~fair election.
Thus, we assume it is true throughout our analysis,
with the exception of Sections \ref{sec35}--\ref{sec37}, where we suppose there were
irregularities in the allocation of voters to notebooks.

{\spaceskip=0.18em plus 0.05em minus 0.02em Next, let us introduce the basic notation used\break throughout} this paper.
To do so, I will use the term \textit{polling unit} generically in the
next three paragraphs to refer to a center or a table or a notebook.
\begin{itemize}
\item[--]
Let $Y_i$ be the number of YES votes (those favoring recalling
President Ch\'avez) and $N_i$ the number of NO votes in polling unit $i$.
\item[--]
Let $T_i = Y_i+N_i $ be the total number of \textit{valid votes} in
polling unit $i$ and $\tau_i$ the number of \textit{voters} assigned to
that polling unit (the \textit{size} of the polling unit).
Note the difference between \textit{voters} and \textit{valid votes}.
\item[--] Let $O_i = \tau_i - T_i$ be the number of invalid votes and
abstentions in the polling unit $i$.
For brevity, we refer to them as the OUT votes (out of the electoral
consultation).
\end{itemize}

In the rest of this section, where we review different papers, the
subscript can refer to centers, tables or notebooks. However, in
Section \ref{sec3} the subscripts are used only to identify voting notebooks.

\subsection{Discrepancies Between Two Exit Polls and Official
Results}\label{sec21}

Prado and Sans\'o (\citeyear{prado}) addressed the controversial
discrepancy between two independent exit polls and the official results.
Roughly, the official result was $41\%$ YES votes and $59\%$ NO votes,
while the exit poll results were $61\%$ YES votes.
The polls were collected by a political party (Primero Justicia) and a
non-governmental organization (S\'umate), both opposition to president
Ch\'avez.
The authors' main claims are:
\begin{enumerate}[C2:]
\item[C1:] There was no selection bias in choosing the centers to be polled.
\item[C2:] The discrepancies per center cannot be explai\-ned by sampling errors.
\end{enumerate}

C1 is settled by noting that the proportion of YES votes for the
overall population matches the proportion of YES votes for the polled centers.

Claim C2 is addressed by assuming that the sampling distribution of the
number of YES \textit{answers} for a given polled center $i$, say $y_i$,
is a Binomial($t_i,p_i$) random variable.
The parameters of this Binomial are: $t_i$ the size of the sample
collected at the center and
$p_i$ the proportion of YES votes, namely $p_i = Y_i/T_i$.
Under this assumption, Prado and Sans\'o (\citeyear{prado}) showed that
there are significant differences between the official results and the
exit polls in about 60\% of the $497$ polled centers.
The authors also considered the pairwise comparison between the two
exit polls among the common polled centers (27 in total). We remark
that eight of them (30\%) differ significantly.

It appears that Prado and Sans\'o had the following assumptions in mind
to determine that $y_i$ is Binomial with the parameters above:
\begin{enumerate}[A2:]
\item[A1:] Given a polling center, the persons to interview were
selected by \textit{simple random sampling.}
\item[A2:] Each interviewed person responded to the question with the truth.
\end{enumerate}

A careful reading of Section 2 of Prado and Sans\'o (\citeyear{prado})
suggests that the sample at each center may correspond to a more
complex model than simple random sampling.
How could the used model affect the estimates and conclusions of their analysis?
If, for example, and as seems to be, \textit{stratified sampling} was
used, it will depend on the \textit{stratification} schema and the \textit
{allocation} criteria used by the pollsters (Lohr, \citeyear{lohr1999}).
In the absence of concrete information, the assumption of the binomial
distribution is the most reasonable one.
However, we cannot ignore the uncertainty about the model and,
consequently, about the sampling errors computed under~A1.

The authors discussed briefly the consequences of the non-veracity of A2.
``It has been demonstrated repeatedly that non-response can have large
effects on the results of a survey'' (Lohr, \citeyear{lohr1999}).
It is quite possible that, in a highly polarized political climate,
voters that supported Ch\'avez were associated with non-response, since
they could identify the pollsters as members of the opposition to Ch\'avez.
Unfortunately Prado and Sans\'o had no estimates of non-responses and
so had to ignore their effects.

Other sources of voter selection bias and measurement error are
discussed in this paper.
Some of them could imply a systematic bias across the pollsters.
Such is the case of the late closing of the voting centers:

\textit{The voting centers had to be open until 4:00 p.m. but the
electoral umpire extended the closing time twice, first until 9:00 p.m.
and finally until midnight.
This was not foreseen by the pollsters and during the afternoon and
evening, there was a fierce campaign to promote the attendance of the
supporters of President Ch\'avez to the voting centers} (The Carter
Center, \citeyear{cc2005}).

Prado and Sans\'o (\citeyear{prado}) also studied this possibility, but
the available data are very limited.
Although the statistical procedure and motive are correct, missing data
can produce results that have no validity at all (De Veaux and Hand,
\citeyear{veaux2005}).

It is hard to believe that the discrepancies between exit polls and
official results are due to sampling and random non-sampling errors.
Unfortunately the information about exit polls is limited and does not
allow a more rigorous analysis.

\subsection{YES Votes Versus Number of Signers in~the~Recall
Petition}\label{sec22}

Delfino and Salas (\citeyear{delfino}) focused on the association
between the YES votes and the number of signers in the recall
petition.\footnote{For readers who do not know the intricacies of the
referendum, the signatures were collected eight months before the
referendum. Many signers were invalidated and some had to sign again in
a second runoff (The Carter Center, \citeyear{cc2005}).}
In the first four sections of this paper the authors described the
electoral process well.
However, from the fifth section onward, I~have major concerns.

Let $S_i$ be the number of signers in voting center~$i$.
The authors considered the following two \textit{relative numbers} of YES
votes and signers:
%
\begin{equation}
k_i = \frac{Y_i}{S_i}\quad \mbox{and} \quad s_i =\frac{ S_i}{T_i}.
\end{equation}
They conducted a bivariate data analysis with $k$ as a response
variable and $s$ as an input variable.
Since $k \leq1/s$, they said: ``In voting centers with a large value
of $s$, we expect a value of $k$ around 1$\ldots$
The situation is completely different in voting centers with a small
value of $s$. The singularity can produce very high values of $k$ in
the neighborhood of $s = 0$. Hence the level of uncertainty in $k$
becomes very large.''
Later on, they added: ``The computerized centers are very far away from
$1/s$, clearly contradicting the expected non-linear behavior with
respect to $s$.''
Finally, they claimed fraud because the data contradict this behavior
and even ventured to establish a hypothesis:
``In computerized centers, official results were forced to follow a
linear relationship with respect to the number of signatures.''

What can justify the previous conjecture?
All that we really know is that the range of $k$ is larger when $s$ decreases.
How can we infer the \textit{expected nonlinear behavior of $k$ with
respect to $s$} from this fact?
As is shown in equation (4) of their paper,
\[
k_i = \frac{p_i}{s_i},
\]
$p_i = Y_i/T_i$ being the proportion of YES votes in center $i$.
Then, $k$ decreases as $1/s$, of course, but increases as $p$ does and
there is a strong relation between these two variables.
In fact, as we will explain next, one expects the value of $k$ to be
constant with respect to $s$, not only showing that the conjecture of
Delfino and Salas (\citeyear{delfino}) is false, but showing that the
results observed are as expected.

Following their schema, we analyze the (full) computerized centers and
(full) manual centers separately.
Manual centers are peculiar. They usually correspond to remote
locations and they have a much smaller number of votes than the
computerized ones (Prado and Sans\'o, \citeyear{prado}).
For this reason many authors perform a separate analysis of these data.
There was also a small number of mixed centers 
where there were both manual and computerized notebooks.
These centers represent only 1.26\% of the total YES votes, 1.3\% of
the valid votes, and are excluded in what follows.

Let $\gamma_1 = 389\mbox{,}862$ and $\gamma_2 = 3\mbox{,}548\mbox{,}811$, the total YES
votes in manual and computerized centers, respectively.
Consider also the total number of signers in manual and computerized
centers, that we shall denote by $\theta_1 \gamma_1$ and $\theta_2
\gamma_2 $,
so that $\theta_1$ and $\theta_2$ are ratios between total signers and
total YES votes.
As mentioned before, I am skeptical about the use of data that are not
official results of the referendum.
So, we will assume $\theta_1$ and $\theta_2$ are unknown parameters and
will only assign values to them for simulation purposes.

%
\begin{figure*}

\includegraphics{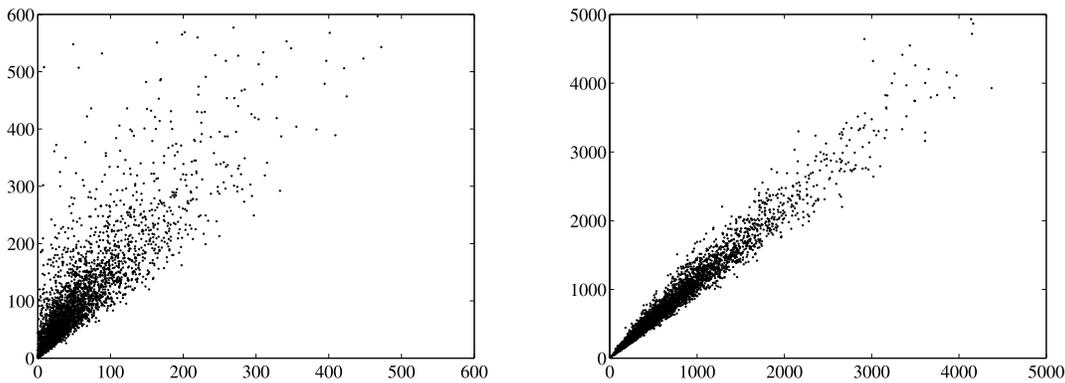}

\caption{YES votes versus simulated signatures according to the
heteroscedastic linear model (\protect\ref{hardcore}). The left panel
corresponds to manual centers with $\theta_1 = 1/1.81$. The right panel
corresponds to computerized centers with $\theta_2 = 1/1.15$.}\label{fig1}
\end{figure*}

As The Carter Center (\citeyear{cc2005}) remarked, the signers were the hard core of the YES votes.
In fact, Delfino and Salas (\citeyear{delfino}) claimed that ``each
signature has a high probability of resulting a YES vote.''
Let us simplify the scenario and assume that each signature in a center
was a YES vote in that center.
Thus, the ratios $\theta_1$ and $\theta_2$ are less than 1.
Under this assumption, the conditional distribution of~$S_i$ given
$Y_i$ can be fitted by a hypergeometric distribution with parameters
$\gamma_c$ (\textit{the number of marbles} in the hypergeometric jargon),
$\theta_c\gamma_c$ (\textit{the number of white marbles}) and $Y_i$
(\textit{the number of draws}), $c$~being equal to $1$ or $2$ according to
whether~$i$ represents a manual or computerized center. The expected
value and variance of the hypergeometric variable are
\begin{eqnarray}
\EE[S_i|Y_i] &= &\theta_cY_i \quad\mbox{and}\nonumber\\ [-8pt]\\ [-8pt]
\operatorname{Var}[S_i|Y_i] &=& Y_i\theta_c(1-\theta_c)\frac{\gamma_c-Y_i}{\gamma_c-1}.\nonumber
\end{eqnarray}
Using the standard normal approximation one obtains
%
\begin{equation}\label{clt}
\frac{S_i-\theta_c Y_i}{\sqrt{Y_i\theta_c(1-\theta_c)}}\approx \mathcal{N}(0,1),
\end{equation}
$\mathcal{N}(\mu,\sigma^2)$ a Normal random variable with mean $\mu$ and
variance $\sigma^2$.
Relation (\ref{clt}) leads us to consider the two heteroscedastic
linear models
%
\begin{equation}\label{hardcore}
S = \theta_c Y +\mathcal{N}\bigl(0,\theta_c (1-\theta_c) Y\bigr),
\end{equation}
for manual centers ($c=1$) and computerized centers ($c=2$).

For each center, we simulated the number of signatures given the number
of YES votes at the center using~(\ref{hardcore}).
Typical outcomes of these simulations with $\theta_1 = 1/1.81$ and
$\theta_2 = 1/1.15$ are shown in Figure \ref{fig1}.
The values of $\theta_c$ were chosen with the intention of comparing
our simulated clouds of points with those shown in Figure 6 of Delfino
and Salas (\citeyear{delfino}).
Note that the least squares regression lines of the latter ones have
slopes $1.81$ and $1.15$ using a~reverse relation between the\vadjust{\goodbreak}
variables, namely $Y = a_c S  +  b_c  +  \mathrm{error}$.
Thus, we take $\theta_c = 1/a_c$.
It is difficult to see how to reject the regression model (\ref{hardcore})
using statistical testing, even under the classical
homoscedastic linear model.
The differences between the clouds associated with manual and
computerized centers are due to differences in scale and variances,
included in the heteroscedastic linear model~(\ref{hardcore}).\break
There is nothing mysterious about this difference, as Delfino and Salas
(\citeyear{delfino}) suggested.
Reversing the relationship between $Y$ and $S$ in regression model~(\ref{hardcore})
yields a heteroscedastic linear model
%
\begin{equation}
Y = \beta_c S + \mathcal{N}(0,\sigma_c ^2 S).
\end{equation}
Dividing by $S $, the above equation becomes
%
\begin{equation}\label{harcore2}
k = \beta_c + \frac{1}{\sqrt{S}} \mathcal{N}(0,\sigma_c^2),
\end{equation}
which precisely describes the clouds of points shown in Figures 3 and 5
of Delfino and Salas (\citeyear{delfino}), with observations around a
constant for any value of $s$, although the range of $k$ is larger when
$s$ is smaller.
In summary, it is expected that $\{k_i\}$ will be constant with a
dispersion which decreases as $1/\sqrt{S}$ (note the difference between
$S=sT$ and $s$).
Note that, although $s$ will be small, if $T$ is large (like almost
every computerized center), the variance can be small, explaining why
computerized centers are more concentrated around the expected value of $k$.

\begin{figure*}

\includegraphics{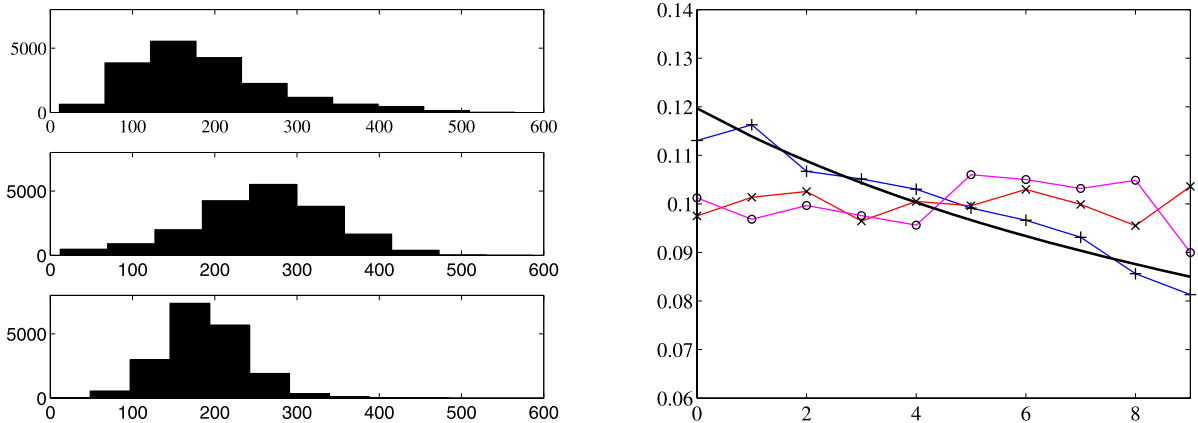}

\caption{Left panel: Histogram of the YES votes (top), NO votes
(middle) and OUT votes (bottom) per notebook. Right panel: Benford's
Law for the second digit (solid line) versus relative frequencies of
the second digit for YES votes $=+$, NO votes $=\times$ and OUT votes
$=\circ$.}\label{fig2}
\end{figure*}

There is an additional comment related to Figures~3, 4 and 5 of Delfino
and Salas (\citeyear{delfino}) worth making.
Note that all right panels have a gap for small values of the input
variables (almost without observations).
Compare the figures removing these gaps in both panels.
For example, remove the windows with $s<0.1$ in Figures 3 and 5 and the
windows with less than 200 total votes in Figure 4.
The behavior is very similar for manual and computerized voting centers.
Their conclusion about the different behavior between manual and
computerized centers seems inaccurate.
%

%
\begin{figure*}

\includegraphics{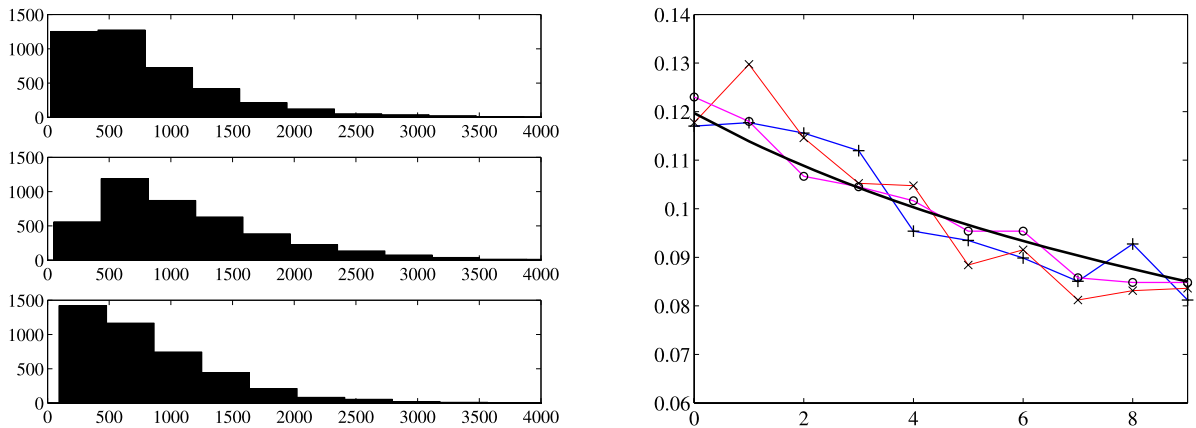}

\caption{Left panel: Histogram of the YES votes (top), NO votes
(middle) and OUT votes (bottom) aggregated by center. Right panel:
Benford's Law for the second digit (solid line) versus relative
frequencies of the second digit for YES votes $=+$, NO votes $=\times$
and OUT votes $=\circ$.}\label{fig3}
\end{figure*}

There are more intriguing statistical arguments in the paper of Delfino
and Salas (\citeyear{delfino}).
Although we have focused only on their main claim, I should add a
comment related to the data.
From the least squares regression lines shown in Figure 6 of Delfino
and Salas (\citeyear{delfino}) one can estimate the total signatures in
fully manual or computerized centers (excluding the mixed ones) on
which the authors base their study.
This total is 3,310,200, close to the\break 3,467,051 signatures submitted to
the electoral umpire (Delfino and Salas, \citeyear{delfino}).
However, the total number of valid signers was 2,553,051 (The Carter
Center, \citeyear{cc2005}).
I~leave the conclusion to the reader.

\subsection{Anomaly Detection by Benford's Law}

Pericchi and Torres (\citeyear{pericchi}) compared empirical
distributions with Benford's Law governing the frequency of the
significant digits (Hill, \citeyear{hill1995}).
Considering several electoral processes in three countries, the only
case compellingly rejected by their test is the NO votes at
computerized notebooks in the Venezuelan recall referendum.
In addition, they made reference to recent contributions in which
compliance or violation of the law in electoral processes has been studied.
Some criticisms related to the use of the law in electoral data (The
Carter Center, \citeyear{cc2005}; Taylor, \citeyear{taylor}) were
also discussed.
As theoretical contributions, the authors obtained a generalization of
the law under restrictions of the maximum number of votes per polling
station and discussed technical issues related to measuring the fit of
the law.


It is important to note that Pericchi and Torres (\citeyear{pericchi})
did not analyze the OUT votes or abstentions. Figure \ref{fig2} shows the
marginal distributions of each option of vote per notebook (left panel)
and compares the empirical distributions of the second digit with
Benford's Law (right panel).
Regarding Figure \ref{fig2}:
\begin{itemize}
\item As Pericchi and Torres showed, the YES votes conform to the law,
while the NO votes do not. However, the strongest widespread departure
from the law is related to the OUT votes. The $\chi^2$ test statistic
for this option is the highest of the three.
\item It is known that compliance with the law is more likely when the
skewness is positive (Wallace, \citeyear{wallace2002}), and the only
distribution with positive skewness is related to the YES votes.
\end{itemize}

We should remark that violations of Benford's Law 
may be due to unbiased errors
(Etteridge and Srivastava, \citeyear{etteridge1999}).
Thus, deviations from the law can arise regardless of whether an
election is fair or not (Deckert et al., \citeyear{deckert}).
On the other hand, there are many types of fraud that cannot be
detected by Benford's analysis (Durtschi et al., \citeyear{cindy2004}).
So, electoral results that conform to the law are not neccessarly free
of suspicion. 

To illustrate the comments above let us consider results by centers
rather than by notebooks. In Figure \ref{fig3} (left panel) we show the
distributions of the number of votes at this aggregation level.
Note that now all distributions have positive skewness.
In the same figure (right panel) we also show Benford's Law for the
second digit and the related empirical distributions of vote per center.
All voting options confirm the law. According to this analysis, there
is no reason to doubt the official results by center, despite that the
test suggests the contrary when we use the results by notebook. Is the
former a false negative or the latter a false positive?
Could unbiased errors in the vote counting by notebooks reproduce such
a scenario?
Or, conversely, could the results by centers be masking a fraud in notebooks?
Benford's test does not address this controversy.\looseness=1

\subsection{Irregularity in the YES Votes Distribution}

Febres and Marquez (\citeyear{febres2006}) tested the distribution of
YES votes in the voting notebooks.
In a~first round, they applied a $Z$ test to compare the proportion of
YES votes in each notebook with the proportion from the center to which
the notebook belongs.
The number of \textit{irregular notebooks} (notebooks with a proportion
significantly different from the proportion of the center) resulting
from this round is expected.
Therefore, this analysis suggests no inconsistency.
According to the territorial organization of Venezuela, the voting
centers are grouped into parishes.
The authors subdivided the parishes into clusters of centers, using a
criterion that we discuss later.
They then applied Pearson's $\chi^2$ test to compare the distribution
of YES votes among the notebooks at each cluster with the conditional
expected distribution given the overall results by cluster and valid
votes by notebook.
In this second round, they reported a high percentage of \textit{irregular
clusters} (clusters with an outlier $\chi^2$ statistic).
Their main finding was that the irregular clusters favor the NO option.
Moreover, they showed a monotone relationship between the proportion of
YES votes by cluster and the $p$-value of the Pearson $\chi^2$ test.
Tuning the confidence level to block irregular clusters, they estimated
the overall result and the winning option is YES.

%
\begin{figure*}

\includegraphics{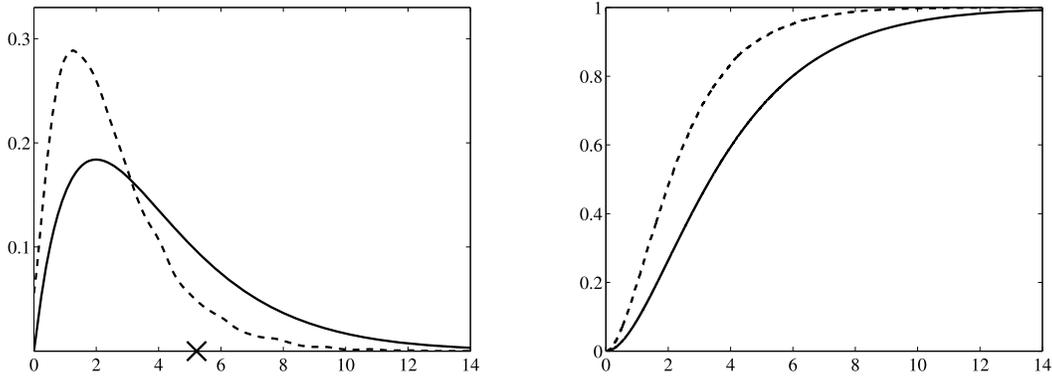}

\caption{Left panel: Exact probability density function (dashed
line) of the $\chi^2$ test statistic related to the cluster with five
notebooks described in Table 9 of Febres and Marquez (\protect\citeyear
{febres2006}). Probability density function of the reference
distribution used by the authors (solid line). The cross marks the
observed value for the test statistic. Right panel: Exact cumulative
distribution function (dashed line) and usual asymptotic approximation
for the $\chi^2$ test statistic (solid line).}\label{fig4}
\end{figure*}

As mentioned earlier, voters within the same center were randomly
assigned to the notebooks.
Thus, each notebook is a random sample without replacement from the
voting center population.
The framework can be completely different when notebooks are grouped by
clusters of centers.
If the proportions of YES votes of two centers in the same cluster are
not equal, no matter how similar they are, and if the total number of
votes by notebooks is large enough, any consistent test will detect
significant discrepancy between the proportions in the notebooks and
the proportion in the cluster.
The authors took care of this fact.
They made a trade-off between the homogeneity of the cluster (how
similar the proportions of the centers within the cluster are) and the
number of votes per notebook.
Basically, the clusters were chosen such that the $Z$ test does not
detect a~significant difference between the proportion of YES votes at
the notebook with the greatest number of votes and the cluster proportion.
In this way they ensured that each notebook is a representative sample
of the cluster.
The authors referred to this as the \textit{minimum heterogeneity
distance for clustering analysis} and made reference to the books of
Sokal and Sneath (\citeyear{sokal1973}) and Press (\citeyear{press1982}).

I have two concerns about these results. The first deals with a general
concern about the validity of ad hoc mechanisms to identify false
positives (detecting fraud when none is present), which might be the
case. The second is a technical issue that must be resolved before
subscribing to the authors' conclusions.

In the referendum context, the standard cluster units of notebooks are
the voting centers.
I guess that the authors did not report results at this level of
aggregation because they did not observe inconsistencies at this level
of aggregation.
In fact, if we apply Pearson's $\chi^2$ test to detect \textit{irregular
centers}, in the same way that the authors applied this test to detect
\textit{irregular clusters}, we do not observe major inconsistencies.
Therefore, their results depend on a particular way of clustering the notebooks.
Why these clusters instead of ones more or less homogeneous?
Why keep the hierarchical ordering by par\-ishes instead of another more
related to political preferences?
With these questions I am only trying to illustrate natural doubts that
can arise when we introduce ad hoc criteria for grouping notebooks.
If the results were independent of the grouping level, then this would
not matter, but this is not the case.

My second concern is the use of the usual asymptotic distribution of
Pearson's $\chi^2$ statistic to determine when an observed value of the
test statistic is an outlier.
This asymptotic does not hold in the framework that we are considering.
In general, it is doubtful that this holds when the multinomial
distribution,\vadjust{\goodbreak} which is the standard underlying assumption when this
test is performed, is replaced by a multivariate hypergeometric
distribution (Zelterman, \citeyear{zelterman2006}), which is the
reference model for the distribution of votes among notebooks.
In particular, because \textit{all} the votes of each cluster are
distributed among the notebooks, the correlations are not negligible.
Despite this, I do not deny that there is a high percentage of
irregular clusters.
To illustrate the previous comment, we consider the cluster with five
notebooks described in Table 9 of Febres and Marquez (\citeyear{febres2006}).
Following the standard asymptotics, the authors used the $\chi^2$
distribution with four degrees of freedom to compute the $p$-value of
the test statistic related with this cluster.
We compute the exact distribution of this statistic to compare with the
$\chi^2(4)$ distribution.
How to compute the exact distribution is not relevant for now (it is a
simple exercise following the discussion in Section \ref{sec3}).
The important thing here is that an outlier for the $\chi^2(4)$
distribution is also an outlier for the exact distribution (see the
left panel of Figure \ref{fig4}).
In fact, as the right panel of Figure \ref{fig4} shows, the test statistic for
this cluster is \textit{less} than $\chi^2(4)$ \textit{in the usual
stochastic order.}
If we had a similar result for all clusters, then we could ensure that
the percentage of irregular clusters is equal to or greater than the
percentage reported in the paper.
I believe that such a result could be obtained.
An alternative would be to compute the exact distribution for each
cluster to recompute the $p$-values and the percentage of outliers.
This involves high computational costs but it would also allow us to
test the authors' main claim about a monotone causal relationship
between the proportion of YES votes and the $p$-value.

The conjectures of Febres and Marquez are interesting and point in a
concrete direction, but require a further analysis before raising them
to conclusions of fraud.

\subsection{Too Many Ties?}

Taylor (\citeyear{taylor}) considered the following six models of
``fair elections'':
\begin{enumerate}[T3.1.]
\item[T1.]
A model in which the YES/NO votes in computerized notebooks are
independent and\break identically distributed Poisson random variables, with
common expectation according to the results in the country.
\item[T2.]
The same model as above but with a common distribution which is not
necessarily Poisson.
\item[T3:]
A model in which the YES/NO votes in the notebooks of each electoral
table are independent and identically distributed Poisson random
variables, with common expectation according to the results in the table.
\item[T3.1.]
A model in which the distribution of YES/NO is multinomial, splitting
up the YES/NO votes of each electoral table equally among the notebooks.
\item[T4.]
A multivariate hypergeometric model, conditioned on the results per
electoral table and valid votes per notebook.
%
\item[T5.]
A parametric bootstrap where total votes of notebooks $\{T_i\}$ are
generated according to the integer part of a multivariate Normal distribu\-tion.
Then YES votes in notebook $i$ are samp\-led according to a
Binomial($T_i,p$), $p$ being the proportion of YES vote in the
electoral table.\looseness=-1
\end{enumerate}
Although in Taylor's paper it is not always explicitly said, T3--T5 are
conditioned on the official results by electoral table and T4 is
additionally conditioned on the official number of valid votes by notebook.

From these models, the author analyzed different statistical anomalies
related with claims of fraud.
Two of them have been previously discussed in this section
(Febres and Marquez \citeyear{febres2006}; Pericchi and Torres \citeyear{pericchi}). The third is related to high rates of YES
ties: A YES tie is a perfect match of YES votes between two notebooks.
Accordingly, his analysis can be divided into three parts:
\begin{itemize}
\item Global test for goodness of fit for models T3\break and~T3.1.
\item Comparative study between the distribution of the significant
digits according to T3.1 (also to a slight improvement of T1), the
observed distribution and Benford's Law.\vadjust{\goodbreak}
\item Computation of the expected number of electoral tables with one
or more YES ties, for each model; and comparison with the observed
number of ties.
\end{itemize}
His main results and conclusions can be summarized as follows:
\begin{enumerate}[R3.]
\item[R1.] ``The more powerful $\chi^2$ test'' strongly rejects the
Poisson model T3. However, a \textit{False Discovery Rates} analysis
(Benjamini and Hochberg, \citeyear{benjamini1995}) shows ``there is
not evidence of widespread departures for the Poisson model.''
This result ``shows no systematic fraud in the form of vote-capping.''
\item[R2.] The distribution of the significant digits of the
multinomial model T3.1 does not conform to Benford's Law and is
virtually identical to the observed distribution.
Thus, Benford's Law is of ``little use in fraud detection in this instance.''
\item[R3.] The $Z$ scores used to compare the observed number of
electoral tables with one or more YES ties with the expected number
according to his models ``are fairly high'' (I will make an exception
with the $Z$ test related to T4, which is equal to 2.37).
``This only means that we can reject the global null hypothesis''
(i.e., the global models) ``and not that there indeed was fraud.''
\end{enumerate}

First of all, the validity of a statistical model is not entirely
justified by the fact that it fits the data, especially if one wants to
test the quality of those data.
The costs, here associated with a false negative (failing to identify a
fraud condition when one exists), are too high.
The model should at least not be at odds with our knowledge about the
system that is being modeled.

According to Taylor's web page,\footnote{\href{http://www-stat.stanford.edu/\textasciitilde jtaylo/venezuela}{http://www-stat.stanford.edu/\textasciitilde jtaylo/venezuela}.}
``the first two
models'' (T1 and T2) ``are clearly unrealistic.''
The next two (T3 and T3.1) also are:
\begin{enumerate}[(b)]
\item[(a)]
As discussed in the previous subsection, the assumption of independence
among the notebooks is meaningless. There are links on the sums of
votes across an electoral table. All the votes
of each table are distributed among its notebooks, so the correlations
are not negligible.
\item[(b)]
The number of voters (note again the difference between voters and
votes) by notebooks varies among the notebooks of the same electoral
table, so it makes no sense to equally split the votes of a table among
the notebooks.\vadjust{\goodbreak}
\end{enumerate}

The last two models (T4 and T5) take into account (a) and (b).
In particular, I agree that the multivariate hypergeometric approach
used in T4 is the right way to generate vote configurations.
However, T5 resorts to assumptions that can be questionable, as to the
use of the integer parts of multivariate Normal random variables to
generate valid votes by notebooks.
Given that the two models provide similar results according to his own
analysis, we will apply the principle of Occam's razor\footnote{\textit
{Entia non sunt multiplicanda praeter necessitatem} (entities must not
be multiplied beyond necessity).} to reduce his list to just one \textit
{realistic} model.

What can we conclude when a questionable dataset does not show evidence
of widespread departures for an unrealistic model?
What if the distributions of the significant digits are similar between
them but differ from Benford's Law?
The conclusions in R1 and R2 are baseless.
We cannot conclude anything useful from these analyses.

Let us move to R3, where he considers the multivariate hypergeometric
model and simulations carried out by Felten et al. (\citeyear{felten2004})
to analyze the YES ties phenomenon.
These simulations show that the number of electoral tables with one or
more ties is high, but not high enough to be considered a sign of fraud
(around 1\% of cases can have an equal or greater number of tables with
YES ties, according to this model).
This part of his analysis did not detect extreme statistical anomalies
that would indicate obvious fraud in the referendum.
Of course, as Felten et al. emphasized, this does not imply the absence
of fraud, either.

\section{Reexamining the Referendum}\label{sec3}

The purpose of this section is to reevaluate the claim of fraud.
An electoral fraud occurs if the results are altered to favor one of
the options.
Having evidence that the changes are enough to overturn the winner, the
outcomes of the referendum should not be recognized.
Moreover, if the handling does not change the winner, but changes the
proportions significantly, it must be considered a fraud.
Electoral results can affect drastically future electoral processes.
In particular, this could have happened during the Venezuelan
parliamentary elections, one year after the referendum,
in which the political parties that supported the YES option withdrew,
claiming the possibility of \textit{new} fraud. 
Also, a tight result can have a different political meaning than an
outcome with a winner by a wide margin,
especially in a recall referendum.
At the end of this section we evaluate the hypothesis of irregularities
in the vote counting to favor significantly the NO option.
We begin by describing the joint probability distribution of results
per notebook, conditioned on the complete set of information of each center.
This corresponds to a multivariate hypergeometric model, similar to
that used in Felten et al. (\citeyear{felten2004}) and Taylor's T5
model (the differences are explained below).
This is a key tool in the hypothesis test methodology that we develop
through this section.

\subsection{Shuffling Voting Cards}

Consider a center with $m$ notebooks, labeled by $1,2,\ldots, m$.
Let $\nu= \sum_{i=1}^m\tau_i$ be the total voters in the center.
Identify each voter by a number in $\{1,2,\ldots, \nu\}$ such that the
first $\tau_1$ voters are in notebook 1, the following $\tau_2$ voters
are in notebook 2, and so on.
In the vote counting, each voter is represented by a voting card
according to her/his electoral option.
It can be YES, NO or OUT.
Let $X_i$ be the voting card of voter $i$.
Then, the vote configuration at the center can be represented by
%
\begin{equation}
\mathcal{X} = \overbrace{(\underbrace{X_1,\ldots, X_{\tau_1}}_{\mbox
{notebook 1}},\ldots, \underbrace{X_{\nu-\tau_m+1},\ldots,X_\nu}_{\mbox
{notebook $m$}})}^{\nu\ \mathrm{voters}}.
\end{equation}

Let $y= \sum_{i=1}^mY_i$ be the total YES votes in the center.
Similarly, let $n= \sum_{i=1}^mN_i$ be the total NO votes.
Then, $\mathcal{X}$ is an outcome of shuffling the voting cards of the center:
%
\begin{eqnarray}
\mathcal{C} = (\overbrace{\underbrace{\mathit{YES},\ldots,\mathit{YES}}_{y\ \mathrm{YES\mbox{'}s}},
\underbrace{\mathit{NO},\ldots,\mathit{NO}}_{n\
\mathrm{NO\mbox{'}s}},\underbrace{\mathit{OUT},\ldots,\mathit{OUT}}_{\nu-y-n\  \mathrm{OUT\mbox{'}s}}}^{\nu\
\mathrm{voters}}\nonumber\\
\end{eqnarray}
That is to say that $\mathcal{X}$ is a permutation of $\mathcal{C}$.

According to the random mechanism used by the electoral umpire to
assign voters to notebooks, given $(y, n, \nu)$, any permutation of
voting cards has the same probability of occurring.
This is the underlying statistical principle shared by Febres and
Marquez (\citeyear{febres2006}), Felten et al. (\citeyear
{felten2004}) and Taylor (\citeyear{taylor}) for testing the
referendum data.
However, these authors do not consider all possible permutations:
\begin{itemize}
\item
The sampling distribution of the test used by Feb\-res and Marquez
(\citeyear{febres2006}) in their first round, where they conditioned
on results by centers and valid votes by notebooks, corresponds to
sampling on the set of outcomes of shuffling YES cards and NO cards in
centers, leaving fixed OUT cards in notebooks.
\item
The samples from the multivariate hypergeometric model considered by
Felten et al. (\citeyear{felten2004}) and Taylor (\citeyear{taylor})
belong to a set of permutations even smaller than the previous one.
They conditioned additionally on the total of YES votes and NO votes by
electoral tables.
That is, they just considered shuffling YES cards and NO cards in
tables, also leaving fixed OUT cards in notebooks.
\end{itemize}
Both approaches fail to consider a large number of equiprobable results
that match the referendum results at the centers.
In this paper, we compute sampling distributions of test statistics
considering all possible permutations of the voting cards at each center.
To simplify the writing, in what follows, we will refer to the result
obtained by shuffling randomly the cards across all centers
as a \textit{random sample of the electoral process}.

\subsection{Statistical Hypothesis of Fair Referenda}\label{sec32}

If we assume that the referendum was properly conducted, the results by
notebook correspond to a~random sample of the electoral process.
Therefore, the hypothesis of a \textit{properly conducted referendum} is
\begin{enumerate}[$\mathcal{H}_0$:]
\item[$\mathcal{H}_0$:] \textit{The votes per notebook correspond to a random
sample of the electoral process}.
\end{enumerate}
%
But, the rejection of $\mathcal{H}_0$ does not imply that the results per
notebook were altered to favor one of the options.
It only implies that there is a significant presence of outliers in the
distribution of votes per notebook.
Innocent irregularities,
as the incorrect allocation of voters to notebooks,
can generate such outliers. 
We consider the most innocent alternative to $\mathcal{H}_0$, assuming
that: (1)~there is a significant presence of outliers in the votes per
notebook, (2) the outliers are the result of neutral irregularities,
and (3) the irregularities affect a random set of notebooks, regardless
of whether they belong to strongholds of the winning option or not.
Therefore, we consider the hypothesis of an \textit{atypical fair
referendum}, namely,
%
\begin{enumerate}[$\mathcal{H}_1$:]
\item[$\mathcal{H}_1$:] \textit{There is a significant presence of outliers
in the votes per notebook that is consequence of innocent
irregularities that affect a random set of notebooks.}
\end{enumerate}
If there is in fact a significant presence of outliers, we can reject
$\mathcal{H}_1$ because: (1) the irregularities are not innocent,
introducing a significant bias in the vote counting, or (2) they affect
mostly notebooks in bastions of one of the options.
Therefore, we have to consider the bizarre, but fair, scenario in which
the irregularities that generate the outliers are neutral, no matter
what, and, for some reason, they affect mostly a set of notebooks that
are in strongholds of one of the options.
Thus, we consider the hypothesis of a \textit{bizarre but fair referendum}:
\begin{enumerate}[$\mathcal{H}_2$:]
\item[$\mathcal{H}_2$:] \textit{The significant presence of outliers in the
votes per notebook is the result of innocent irregularities that affect
mostly a set of notebooks from strongholds of one electoral option.}
\end{enumerate}
The remaining alternative is a clear signal that the irregularities are
not innocent. 

Before testing the hypotheses, we describe the
data\-set.

\subsection{Description of the Dataset}

It is required to have at least two notebooks per center for shuffling
voting cards, so we restrict our analysis to these centers.
In addition, since all allegations of fraud are related to computerized
notebooks,
we only consider full computerized centers (centers where there are no
manual votes).
We also exclude a very small number of centers with empty notebooks
(notebooks without valid votes).
Empty notebooks could arise for technical problems, affecting the
distribution of voters to notebooks in such centers.
After this simple cleaning on full computerized centers with two or
more notebooks, a consistent dataset is obtained with 4,162 centers, all
of them with comparable notebooks.
This means that the votes among the notebooks of a center are in the
same order.
These 4,162 centers represent 18,297 notebooks, more than 83\% of the
total voters, and here will be the base of the study.
The mean and the standard deviation of the number of voters per unit
polls are 634 and 73, respectively.

For the last two subsections of this section, we will also use the
results of presidential elections of 1998,
in which Chav\'ez was elected to his first term as President of
Venezuela with 56\% of valid votes against a~coalition of roughly the
same political parties that supported the recall.
This election was carried out with an automated voting system, which
featured a~single integrated electronic network to transmit the results
from the polling stations  to central\vadjust{\goodbreak} headquarters (McCoy, \citeyear{mccoy1999}).
The legitimacy of the electoral process and the acceptance of the
results by political parties and international observers is a guarantee
of the reliability of the results.
At that time, in each center, voters were also randomly assigned to
polling units, according to the last number of their ID,
similarly to the process described in Section \ref{sec2}.
As we do with the referendum's dataset,
we exclude centers with only one unit poll and those with empty unit polls.
After the cleaning, a dataset is obtained with 3952 centers, 15,667
unit polls, that represent 85\% of the total voters.
The mean and the standard deviation of the number of voters per unit
poll are 594 and 112.
For all the above, both sets of data are comparable for the statistical
purposes of Section~\ref{sec36}.

A different scenario overshadowed the presidential elections of 2000,
that we consider in Section \ref{sec37},
in which Chav\'ez was elected to his second term with 59\% of valid votes.
After two years of important political changes, including the enacting
of a new constitution, the criticism of Chavez's government increased,
polarizing the political climate.
Many claims of fraud, including machines not properly functioning,
people whose names did not appear on the electoral registry and
pre-marked ballots, were made at that time.
While The Carter Center does not believe that the election
irregularities would have changed the presidential results, they
consider those elections as flawed and not fully successful (Neuman and
McCoy, \citeyear{cc2000}).
The election was carried out, roughly, with the same voting system used
in 1998.
However, there was an important difference in the number of voters per
unit poll,
increasing significantly the number of centers with only one unit.
As with the referendum and the presidential elections of 1998,
we exclude centers with only one unit poll and those with empty unit polls.
In addition, we exclude unit polls with more votes than voters.
Thus, we obtain a dataset with 1,600 centers, only 3,730 unit polls, that
represents 53\% of the total voters.

The three datasets provide estimates of high precision for the
resultant percentage of votes per electoral option. Table \ref{tab1} summarizes
their main statistics.

\begin{table}
\caption{Statistical summary of the dataset}\label{tab1}
\begin{tabular*}{\columnwidth}{@{\extracolsep{\fill}}lcccc@{}}
\hline
\textbf{Year} & \textbf{Unit polls} & \textbf{\% of total} & \textbf{Mean of voters} &
\textbf{Standard} \\
& & \textbf{voters} & \textbf{per unit poll}& \textbf{deviation}\\
\hline
1998 & 15,667 & 85\% &  \phantom{0}594.70 & 111.97\\
2000 & \phantom{0}3,730 & 53\% & 1662.50 & 361.74\\
2004 & 18,297 & 83\% &  \phantom{0}634.60 & \phantom{0}73.86\\
\hline
\end{tabular*}
\end{table}

%
\begin{figure*}

\includegraphics{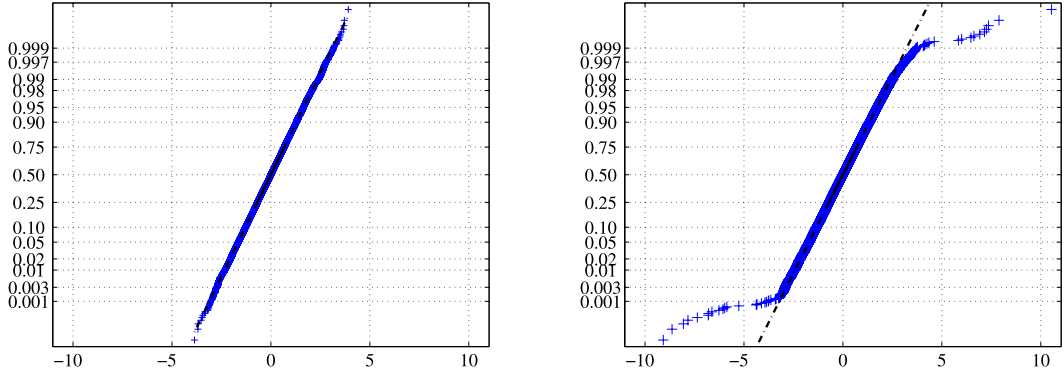}

\caption{Normal probability plot of $Z$ scores based on a random
sample (left) and observed values (right).}\label{fig5}
\end{figure*}

\subsection{Testing the Hypothesis of a Properly Conducted Referendum}


The way to test irregularity is to determine whether an observed value
is an outlier or not.
Let $i$ be a focal notebook and $c$ the center to which it
belongs:\looseness=1
\begin{itemize}
\item[--]
Since voters of a center are randomly assigned to notebooks, $Y_i$ is
the total of YES cards in a simple random sample (without replacement)
of size $\tau_i$ from the voting cards of the center.
In particular, $\EE[Y_i|\mathcal{H}_0] = p_c \tau_i$, with $p_c = y_c/\nu
_c$ and
\[
\operatorname{Var}[Y_i|\mathcal{H}_0] = \tau_i p_c(1-p_c) \frac{\nu_c-\tau_i}{\nu_c -1}.
\]
\item[--]
The minimum $\tau_i$ in the 18,297 notebooks involved is 347 (the mean
value $\bar{\tau} = \sum\tau_i/18{,}297$ is equal to 634.60, and the
maximum is 975).
\end{itemize}
Coupling these facts, a straightforward application of the Central
Limit Theorem implies that, under~$\mathcal{H}_0$, the score
%
\begin{equation}\label{chi1}
Z_i= \frac{(Y_i - p_c\tau_i)}{\sqrt{p_c(1-p_c)\tau_i (\nu_c-\tau
_i)/(\nu_c -1)}}
\end{equation}
is approximately $\mathcal{N}(0,1)$, for any $i$.
Therefore, a test of regularity for a single notebook is reduced to
determining the significance of $Z$.

To get an overall qualitative idea of the joint behavior of the $Z$
scores under $\mathcal{H}_0$, 
the normal probability plot of these statistics from a random sample of
the electoral process is shown in Figure \ref{fig5} (left panel).
In the same figure (right panel) the normal probability plot of the
scores based on the observed values is also shown.
This plot highlights many official results far from what is expected.
Let us peer into the most atypical cases.
Table \ref{tab2} shows the official results of centers 7990 and
 1123,\footnote{We use the center encoding used for the referendum. Codes, as well as
the list of centers, varies from election to election.} where are the
notebooks associated with minimum ($-9.08$) and maximum (10.54) $Z$
value. Let us call these notebooks $m$ and $M$, respectively.
Under $\mathcal{H}_0$, the expected value of $Y_m$ is 161.81, almost twice
the observed value, which is 81, while the expected value of $Y_M$ is
139.08, just over half of what is observed, which is 233.

%
\begin{table}
\tabcolsep=0pt
\caption{Results in centers with notebooks associated with minimum (*)
and maximum (+) $Z$ value}\label{tab2}
\begin{tabular*}{\columnwidth}{@{\extracolsep{\fill}}lccccccc@{}}
\hline
& \multicolumn{3}{c}{\textbf{Center 7990}}&  \multicolumn{4}{c@{}}{\textbf{Center 1123}}\\
\ccline{2-4, 5-8}
\textbf{Notebook} & \multicolumn{3}{c}{$\bolds{m}$} & \multicolumn{4}{c@{}}{$\bolds{M}$} \\
\hline
$Y$ & 174 & \phantom{0}$81^{*}$ & 235 & 191 & \phantom{0}60 & $233^{+}$ & \phantom{0}62\\
$N$ & 272 & \phantom{0}70\phantom{\tsup{*}} & 375 & 396 & 137 & 359\phantom{\tsup{+}} & 143\\
$\tau$ & 607 & 600\phantom{\tsup{*}} & 610 & 588 & 583 & 594\phantom{\tsup{+}} & 567\\
\hline
\end{tabular*}
\end{table}
%


An overall comparison is handled by summing\break squares of $Z$ scores.
Let
%
\begin{equation}
S^2 = \sum_{i=1}^{18{,}297} Z_i^2. 
\end{equation}
A straightforward computation gives $\EE[S^2|\mathcal{H}_0] = \break 18{,}297$.
The variance can be estimated by Monte Carlo, shuffling the voting cards.
We performed 1,000 random samples of the electoral process and obtained
a standard deviation of 216.
Next we show that the sampling distribution of the test statistic
%
\begin{equation}
\label{chi3}
T^{\mathrm{YES}} = \frac{S^2 - 18\mbox{,}297}{216}
\end{equation}
can be approximated by a standard Normal distribution.

The centers have between 2 and 18 notebooks.
The distribution of the centers according to the number of notebooks is
shown in Table \ref{tab3}.

%
\begin{table*}
\tabcolsep=0pt
\caption{Number of clean and fully computerized centers with $m$ notebooks}\label{tab3}
\begin{tabular*}{\textwidth}{@{\extracolsep{\fill}}lccccccccccccccccc@{}}
\hline
$\bolds{m}$ & \textbf{2} & \textbf{3} & \textbf{4} & \textbf{5} & \textbf{6 }& \textbf{7} & \textbf{8} & \textbf{9} & \textbf{10}& \textbf{11} &\textbf{12} &\textbf{13} &\textbf{14} &\textbf{15} &\textbf{16} & \textbf{17}&\textbf{18} \\
\hline
$C_m$ & 1,044 & 820 & 665 & 496 & 380 & 300 & 208 & 110 &54 & 41 & 19 & 12 & 4 & 4 & 2 & 2 & 1\\
\hline
\end{tabular*}
\vspace*{-6pt}
\end{table*}
%
\begin{table*}
\tabcolsep=0pt
\caption{Degrees of freedom (df) related with $\chi^2_{\mathrm{nb(i)}}$}\label{tab4}
\begin{tabular*}{\textwidth}{@{\extracolsep{\fill}}lcccccccccccccccccc@{}}
\hline
$\bolds{i}$& \textbf{1} & \textbf{2} & \textbf{3} & \textbf{4} & \textbf{5} & \textbf{6 }& \textbf{7} & \textbf{8} & \textbf{9} & \textbf{10}& \textbf{11}
&\textbf{12} &\textbf{13} &\textbf{14} &\textbf{15} &\textbf{16} &
\textbf{17}&\textbf{18}\\
\hline
df &4,162& 4,162& 3,118& 2,298&1,633&1,137& 757&457& 249 & 139 & 85 & 44& 25& 13& 9& 5& 3&
1\\
\hline
\end{tabular*}
\end{table*}

The sum of squares can be decomposed as follows:
\[
S^2 = \sum_{i=1}^{18,297} Z_i^2 = \chi^2_{\mathrm{nb(1)}} + \chi^2_{\mathrm{nb(2)}} + \cdots+\chi^2_{\mathrm{nb(18)}},
\]
$ \chi^2_{\mathrm{nb(i)}}$ being the sum along all the centers of the
squares of the $Z$ scores related to the $i$th notebook of a
center.\vadjust{\goodbreak}
Although the results of notebooks belonging to the same center are
correlated, given $\mathcal{H}_0$ they are independent of results in other centers.
In turn, each $\chi^2_{\mathrm{nb(i)}}$ is the sum of independent random
variables and each one is approximately the square of a standard normal
random variable.
Then, $\chi^2_{\mathrm{nb(i)}}$ is approximately $\chi^2$ with $\sum
_{m\geq i} C_m$ degrees of freedom, $C_m$ being the number of centers
with $m$ notebooks.
Table \ref{tab4} lists the degrees of freedom related to $\{\chi^2_{\mathrm{nb(i)}}, 1\leq i\leq18\} $.

\begin{figure}

\includegraphics{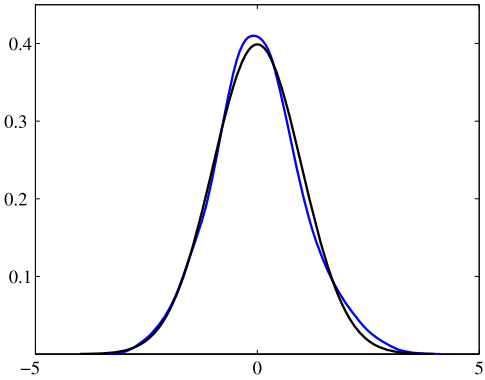}

\caption{Kernel estimator of the probability density function of
$T^{\mathrm{YES}}$ versus a standard Normal probability density.}\label{fig6}
\end{figure}

In general, approximating the distribution of sums of correlated $\chi
^2$ can be difficult.
Fortunately, this is not case here. Two remarks:
\begin{itemize}
\item[--]
For $i\leq10$, the degrees of freedom are large enough to fit the
distribution of $\chi^2_{\mathrm{nb(i)}}$ by a Normal.
\item[--]
$\chi^2_{\mathrm{nb(1)}}+\cdots+\chi^2_{\mathrm{nb(10)}}$ represents 99\% of
the $Z^2$ statistics in $S^2$.
\end{itemize}
Therefore, $S^2$ is approximately a sum of Normal random variables.
Letting $\varsigma^2$ be the sample variance obtained from $k$
independent samples of $S^2$, under $\mathcal{H}_0$, the test statistic
%
\begin{equation}\label{chi3}
T^{\mathrm{YES}} = \frac{S^2 - \EE[S^2|\mathcal{H}_0]}{\varsigma}
\end{equation}
is approximately $\mathcal{N}(0,1)$, for any large $k$.
As we said above, we simulated 1,000 random samples of the electoral
process, obtaining $\zeta=216$.
We also used the samples to confirm that, under $\mathcal{H}_0$, the
distribution of $T^{\mathrm{YES}} $ is approximately $\mathcal{N}(0,1)$.
For that, we test normality with different methods, all of them with
the same conclusive results.
To illustrate, Figure \ref{fig6} compares the kernel density estimator of the
probability density function of $T^{\mathrm{YES}}$ with the probability
density of a standard Normal.

The $T^{\mathrm{YES}}$ observed value, according to the official results, is
$T^{\mathrm{YES}}_{\mathrm{obs}} \approx13.12$, which establishes that the
results of\vadjust{\goodbreak} YES votes per notebook are not credible, given the results
by centers.
The $p$-value, less than the \mbox{MatLab} precision, is strong evidence
against $\mathcal{H}_0$.

%

\begin{figure*}

\includegraphics{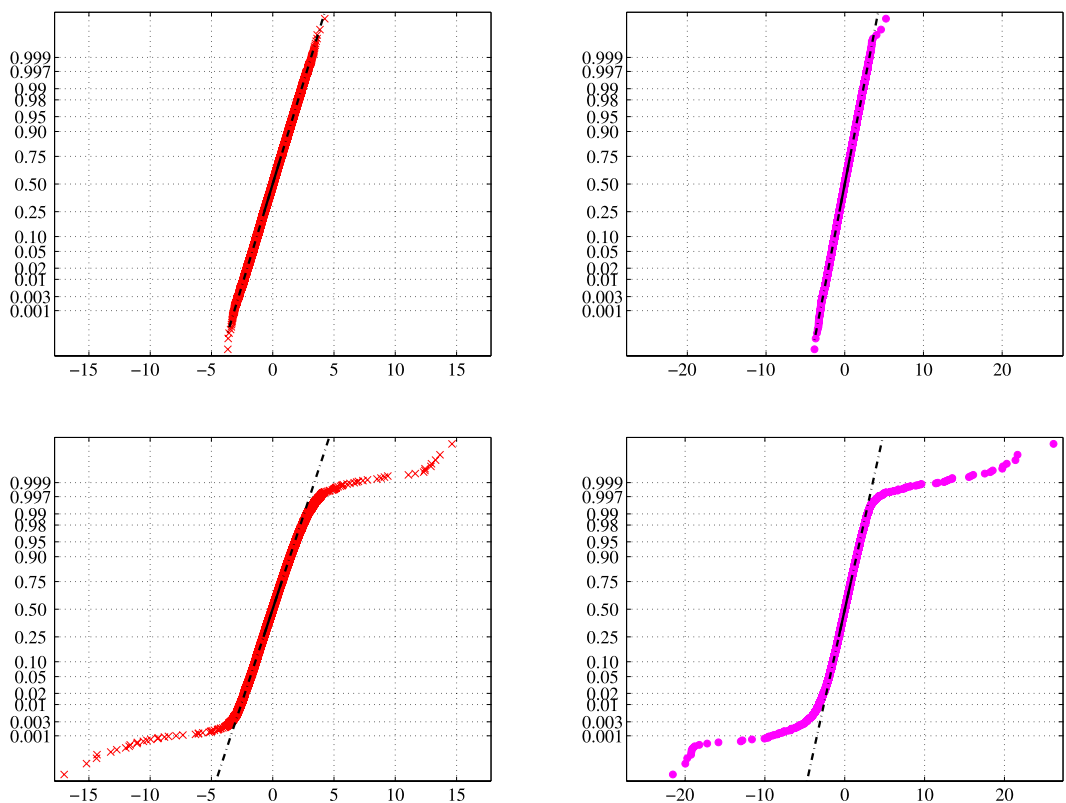}

\caption{Normal probability plot of $Z^{\mathrm{NO}}$ (left) and $Z^{\mathrm{OUT}}$
(right) scores based on a random sample (top) and observed values
(bottom).}\label{fig7}
\end{figure*}

Following the same approach, we can test regularity on the distribution
of NO votes and abstentions.
For that, we define the $Z$ statistics
\begin{eqnarray}\label{ZNO}
\hspace*{25pt}Z_i^{\mathrm{NO}}&=& \frac{(N_i - q_c\tau_i)}{\sqrt{q_c(1-q_c)\tau_i (\nu
_c-\tau_i)/(\nu_c -\tau_i)}}
\quad\mbox{and}\nonumber\\ [-8pt]\\ [-8pt]
\hspace*{25pt}Z_i^{\mathrm{OUT}}&=& \frac{(O_i - r_c\tau_i)}{\sqrt{r_c(1-r_c)\tau_i (\nu
_c-\tau_i)/(\nu_c -\tau_i)}},\nonumber
\end{eqnarray}
$q_c = n_c/\nu_c$ and $r_c = (\nu_c-y_c-n_c)/\nu_c$ being the
proportion of NO votes and OUT votes in the center~$c$ to which
notebook $i$ belongs.
As an illustration, Fi\-gure \ref{fig7} shows the normal probability plots of
these~$Z$ statistics based on a random sample of a properly conducted
referendum.
The figure shows also the normal probability plots of the scores based
on the official results.
These plots show a widespread departure from the expected values, even
stronger than for the YES case (be careful with the scales of these figures).
In fact, if we define test statistics to the distribution of NO
votes\vadjust{\goodbreak}
and OUT votes, $T^{\mathrm{NO}}$ and $T^{\mathrm{OUT}}$ respectively, similar
to what we did for the YES votes, then we have
\[
T^{\mathrm{OUT}}_{\mathrm{obs}}>T^{\mathrm{NO}}_{\mathrm{obs}}>T^{\mathrm{YES}}_{\mathrm{obs}}.
\]
Clearly, $\mathcal{H}_0$ can be completely rejected.

%


\subsection{Testing the Hypothesis of an Atypical Fair~Referendum}\label{sec35}

As mentioned previously, the widespread departure of YESs, NOs and
OUTs per notebook from their expected values could be the outcome of
innocent irregularities in the conduct of the referendum.
Incorrect allocation of voters to notebooks and the passing of votes
from one notebook to another during the vote counting,
by bugs in the programming, are examples of such irregularities.
These irregularities may generate, in particular, $Z^{\mathrm{OUT}}$
outliers but, by the secrecy of the ballot, they should not be
associated with a trend in the vote counting. Next, we propose a
testing methodology, based on a simple statistical control chart, for
testing trend in the vote counting on potentially irregular notebooks.
The methodology can be easily extended to other electoral audit frameworks.
It relies on the assumption that unexpected irregularities can occur in
any unit poll with the same probability.

Denote by $R$ the ratio between NO votes and total valid votes in the
target population, consisting of $K=18{,}297$ notebooks, namely,
%
\begin{equation}\label{pr}
R = \frac{\sum_{i=1}^K N_i}{\sum_{i=1}^K T_i}.
\end{equation}
In sampling jargon, $R$ is the \textit{population ratio} and $K$ is the
\textit{size of the population}.
Let $\mathcal{S}_k$ be the sample consisting of the $k$ notebooks with the
most extreme $Z^{\mathrm{OUT}}$ values.
This is the set of $k$ notebooks with $Z^{\mathrm{OUT}}$ values
furthest
away from zero.
Given a confidence level $1-\alpha$, there is\vadjust{\goodbreak} a $k:=k(\alpha)$ such
that $\mathcal{S}_k$ matches the set of notebooks with $Z^{\mathrm{OUT}}$
values that we consider that are outliers,
that is, the set of notebooks with $Z^{\mathrm{OUT}}$ values out of the
$(1-\alpha)\times100\%$ normal confidence interval.
In our case study, if the confidence level is 99\%, then $k= 706$.
Roughly, 4\% of the $Z^{\mathrm{OUT}}$ values are out of the 99\% confidence
interval.
In what follows, $k$ varies in a range such that $\mathcal{S}_k$
corresponds to the set of outliers, according to some reasonable
confidence level.

Denote by $r_k$ the \textit{sample ratio} based on $\mathcal{S}_k$. That is,
%
\begin{equation}
\label{sr}
r_k= \frac{\sum_{i\in\mathcal{S}_k} N_i}{\sum_{i\in\mathcal{S}_k} T_i}.
\end{equation}
Note that $r_k$ is not the usual ratio estimator, since we are sampling
notebooks with atypical $Z^{\mathrm{OUT}}$ values.
Thus, we might expect that observations $(N_i, T_i)$ in $\mathcal{S}_k$
are larger or smaller than those from a simple random sample (SRS).
However, if the irregularities are innocent, if they do not introduce
bias in the vote counting,
$r_k$ should be similar to the sample ratio based on a SRS.
In particular, if $k$ is large, the bias of the estimator will be small
and the variance can be approximated by
\[
\operatorname{Var}(r_k) \approx S^2_k := \biggl(1- \frac{k}{K}\biggr)\frac{1}{\mu
^2_T}\frac{s^2_r}{k}
\]
(Lohr, \citeyear{lohr1999}), with
\[
\mu_T = \frac{1}{K}\sum_{i=1}^K T_i \quad \mbox{and}\quad s^2_r = \frac
{1}{k-1}\sum_{i\in\textit{S}_k}(N_i- r_k T_i)^2.
\]
Thus, under the hypothesis $\mathcal{H}_1$ defined in Section~\ref{sec32}, if $k$
and $K-k$ are large enough,
%
\begin{equation}\label{score1}
\zeta_k = \frac{r_k - R}{S^2_k}
\end{equation}
is distributed approximately as a standard normal variable. In what
follows, we will only consider $100\leq k\ll K$.

%
\begin{figure*}

\includegraphics{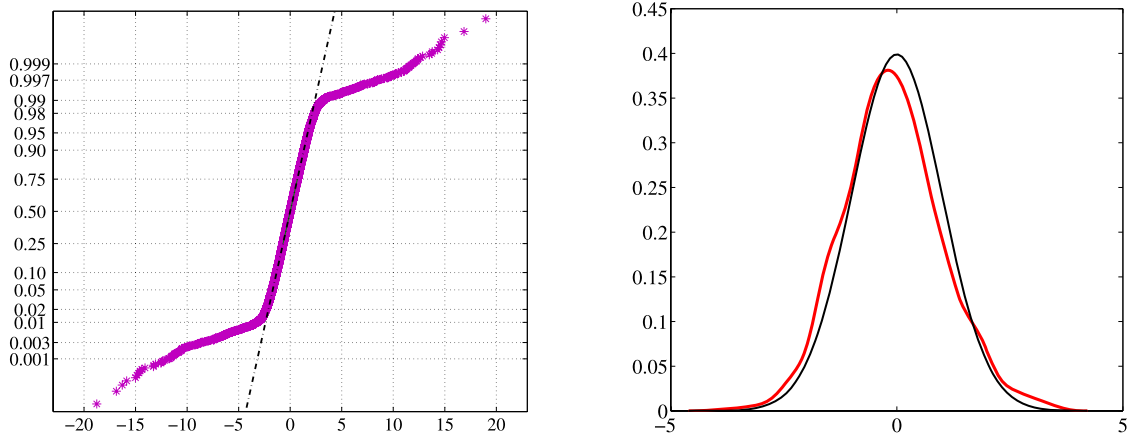}

\caption{Left panel: Normal probability plot of $Z^{\mathrm{OUT}}$ scores from
an atypical fair referendum. Right panel: Standard normal probability
density versus kernel estimator of the probability density function of
$r_{500}$, based on 1,000 independent copies of an atypical fair
referendum.}\label{fig8}
\end{figure*}

To illustrate the above, consider 1,000 independent copies of $\zeta
_{500}$ from a random sample of atypical fair referenda.
We simulate an atypical fair referendum by introducing 700 innocent
irregularities on a~random sample of the electoral process.
Each irregularity consists in passing a random proportion of votes
(10\% on average) from a notebook to another located in the same center.
This handling produces a significant number of $Z^{\mathrm{OUT}}$ outliers
(outside the 99\% normal confidence interval) to those already obtained
before the manipulation.
The normal probability plot of the $Z^{\mathrm{OUT}}$ scores of one of these
atypical fair referenda is shown in Figure \ref{fig8} (left panel) as an example.
As we can see, the shape of the plot is similar to that observed for
the referendum (Figure \ref{fig7}, right bottom panel).
We test normality of $\zeta_{500}$ with different methods, all of them
with the same conclusive positive results. To illustrate, the right
panel of Figure \ref{fig8} compares the kernel density estimator of the
probability density function of $\zeta_{500}$ with the probability
density of a standard Normal.

We can test the hypothesis of an irregular fair referendum using the
$\zeta_k$ scores.
High values of $\zeta_k$ imply that irregularities introduce a bias in
favor of the NO option in the vote counting.
Small values of this score imply a bias in favor of the YES option.
Under~$\mathcal{H}_1$, we expect $\zeta_k$ to be within a confidence
interval.

%
\begin{figure}

\includegraphics{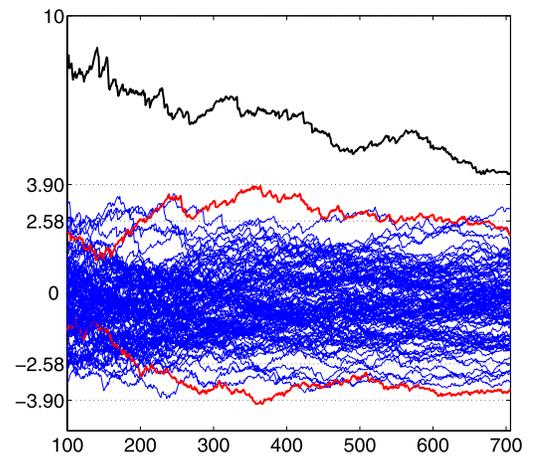}

\caption{$\zeta_k$ versus $k$ for official results (top line) and
simulated atypical fair referenda.}\label{fig9}
\vspace*{3pt}
\end{figure}

The $\zeta_k$ scores corresponding to the official results are plotted
in Figure \ref{fig9} (top line), for $k$ between 100 and 706.
To illustrate the behavior that we expect in an atypical fair
referendum, we plot, in the same figure, 100 simulated scores series of
the atypical fair referendum discussed above.
The $\zeta_k$ scores corresponding to the official results are well
above the 99.99\% confidence interval $(-3.9, 3.9)$ for $100\leq k\leq706$.
Although a small number of simulated trajectories also reach high values,
all of them are embedded in
$(-3.9, 3.9)$ and most of them are in the 99\% confidence
interval\vadjust{\goodbreak}
$(-2.58, 2.58)$, as one expects.
We observed similar behavior in 1,000 additional simulated trajectories
(not plotted).
The scores series of the referendum reaches values higher than any that
we observed in simulations,
being the only one always well above 3.9, for $100\leq k\leq706$.
This provides strong evidence against $\mathcal{H}_1$ than a fairly small
$p$-value of a $\zeta_k$ score, for some $k$.
We are seeing a~significant bias in the vote counting on notebooks
associated with irregularities, which is almost impossible to observe
under $\mathcal{H}_1$.
All the above is strong evidence for rejecting it. 

\subsection{Testing the Hypothesis of Bizarre but Fair
Referendum}\label{sec36}

Most political scientists expect more innocent administrative errors in
areas with more poor voters (M. Lindeman,\vadjust{\goodbreak} personal communication, July 2010).
In addition, ``the conventional wisdom about contemporary Venezuelan
politics is that class voting has become commonplace, with the poor
doggedly supporting Hugo Ch\'avez while the rich oppose him'' (Lupu,
\citeyear{lupu}).
If both beliefs are true, we expect more innocent irregularities in
strongholds of the NO option,
which would explain the atypical result observed in the above section.
That is what~$\mathcal{H}_2$ describes, a general scenario in which
there are more innocent irregularities in centers that support the
winning option.
To illustrate this possibility, we show in Table \ref{tab5} the results in
Center 1123 (C.~M.~A.~Dr. Angel Vicente Ochoa, in Santa Rosal\'ia, Caracas),
one of the most extreme results. All its notebooks are associated with
very extreme $Z^{\mathrm{OUT}}$ values\break (greater than 18.53!).
But the overall NO proportion (65\%) is even less than that observed in
the presidential elections of 1998 (67\%).
This center appears to be a bona fide Ch\'avez stronghold. However, we
have to remark that, in this election, the $Z^{\mathrm{OUT}}$ values of that
center are quite normal, all of them between $-0.4$ and 0.40. The
results, only three unit polls in that election, are shown in Table \ref{tab6}.

%
\begin{table}
\caption{Results in Center 1123 (C. M. A. Dr. Angel Vicente Ochoa, in
Santa Rosal\'ia, Caracas)}\label{tab5}
\begin{tabular*}{\columnwidth}{@{\extracolsep{\fill}}lcccc@{}}
\hline
\textbf{Notebook} & \textbf{1} & \textbf{2} & \textbf{3} & \textbf{4} \\
\hline
$Y$ & 191 & \phantom{0}60 & 233 & \phantom{0}62 \\
$N$ & 396 & 137 & 359 & 143 \\
$\tau$ & 588 & 583 & 594 & 567 \\
\hline
\end{tabular*}
\end{table}

\begin{figure*}

\includegraphics{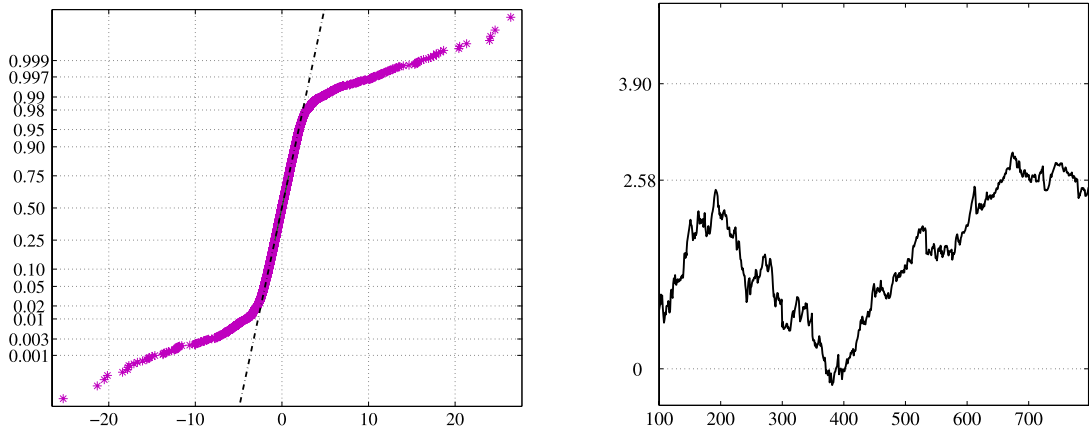}

\caption{Left panel: Normal probability plot of $Z^{\mathrm{OUT}}$ scores of
the presidential elections of 1998. Right panel: $\zeta_k$ versus $k$
for presidential elections of 1998.}\label{fig10}
\end{figure*}

A naive procedure to see if the irregularities affected mostly
notebooks in Ch\'avez's strongholds\break would be
repeating the previous analysis on all centers with outliers.
We consider an alternative analysis for an important reason:
If indeed tampering occurred, it is possible that, in order to mask the
stuff, the irregularities were committed precisely in Ch\'avez's bastions.
In addition, we have to admit that we do not have access to the
re-coding of centers to automatize the procedure.

\begin{table}
\caption{Presidential elections of 1998, results in C.~M.~A.~Dr.~Angel~Vicente Ochoa Center}\label{tab6}
\begin{tabular*}{\columnwidth}{@{\extracolsep{\fill}}lccc@{}}
\hline
\textbf{Unit poll} & \textbf{1} & \textbf{2} & \textbf{3} \\
\hline
$Y$ & 182 & 115 & 123 \\
$N$ & 357 & 265 & 247 \\
$\tau$ & 899 & 645 & 624 \\
\hline
\end{tabular*}
\end{table}

Lupu (\citeyear{lupu}) provided evidence that the presidential election
of 1998 was more monotonic in class voting than the referendum.
This means, the poor were more likely to vote for Ch\'avez in 1998 than
in 2004.
Thus, we expect more innocent irregularities in Ch\'avez's strongholds
in 1998 than in 2004.
In addition, there is not doubt about the legitimacy of this election
(Neuman and McCoy, \citeyear{cc2000}).
For these reasons, the election of 1998 is very appropriate to test if
irregularities affect mostly notebooks in centers that support Ch\'avez,
that is, to test $\mathcal{H}_2$,
defined in Section \ref{sec32}.
The testing schema we use is
to reject $\mathcal{H}_2$ if we fail to reject $\mathcal{H}_1$ for the
elections of 1998.
We begin verifying that there is a significant presence of $Z^{\mathrm{OUT}}$ outliers in 1998:
5\% of $Z^{\mathrm{OUT}}$ values (797 of 15,667) are out of the 99\% normal
confidence interval.
The evidence against $\mathcal{H}_0$ is of the same order as in 2004.
Furthermore, the most extreme $Z^{\mathrm{OUT}}$ values of 1998 are higher
than those observed in 2004.
We omit the details and summarize results by showing the
normal probability plot of the $Z^{\mathrm{OUT}}$ scores in Figure \ref{fig10} (left panel).
It seems possible that, in complex elections, ad hoc decisions are made
to resolve problems that arise on the fly.
As we have discussed previously, this can produce large outliers in the
vote distribution.
However, the test discussed in the previous section strongly supports
$\mathcal{H}_1$ for the presidential elections of 1998.
The corresponding scores series $\{\zeta_k, 100\leq k\leq797\}$ is
almost embedded in the 99\% confidence interval; see right panel of
Figure \ref{fig10}.
%
Therefore, we see that there is little reason to think that
the significant presence of $Z^{\mathrm{OUT}}$ outliers,
that are the result of innocent irregularities, affect mostly a set of
notebooks from Ch\'avez's strongholds.
Irregularities seem to occur randomly, regardless of whether the
notebook belongs to a Ch\'avez bastion or not, and
thus, we reject~$\mathcal{H}_2$.

%


\subsection{Estimating the Effect of the Irregularities}\label{sec37}

We have provided statistical evidence that there was a significant
presence of irregularities that favored the winning option in the vote
counting of 2004. But, how much could the irregularities affect the
overall\vspace*{1pt} results?
Suppose that the $Z^{\mathrm{OUT}}$ outliers are the \textit{tip of the
iceberg} and there is bias in the vote counting of a high proportion of
notebooks, not just in the notebooks with extreme $Z^{\mathrm{OUT}}$ values.
To evaluate this assumption,
we analyze the behavior of the sample ratio\vadjust{\goodbreak} $r_k$ in (\ref{sr}) for
higher values of~$k$ than those that we have already considered.
Figure~\ref{fig11} shows that proportion for $k$ varying from 100 to 3000 (top line).
The shape shows a strong correlation between the trend in the vote
counting and the discrepancy between valid votes and its expectation,
not only in notebooks with $Z^{\mathrm{OUT}}$ outliers.
We remark that for $k=100$ we are considering 41,533 valid votes, a
very large sample size for estimating proportions (an accepted standard
for pollsters is above 1,000).
What we expect when we increase the sample size is exactly what we have
for the presidential elections of 1998 (line from the middle to the
bottom in Figure \ref{fig11}):
%
\begin{figure}

\includegraphics{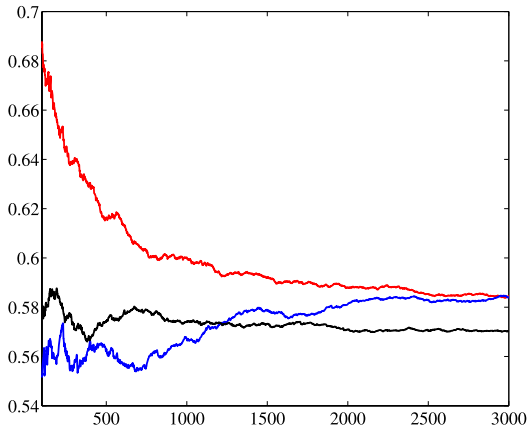}

\caption{Proportion of Ch\'avez's votes in $\mathcal{S}_k$ versus $k$
for the referendum (top line) and presidential elections of 1998 (line
from the middle to the bottom) and 2000 (line from the bottom to the
middle).}\label{fig11}
\end{figure}
The proportion is a function of the sample size that slightly varies
around the population proportion, and that quickly stabilizes around
this value.
So, our assumption of irregularities that affect the vote counting
across all the notebooks is quite possibly true.\footnote{Mart\'in
(\citeyear{isbelia}), which unfortunately was not available for my
review, studies the volume of traffic in incoming and outgoing data
between notebooks and totalizing servers.
It provides evidence that the vote counting of a high percent of
notebooks could be affected from the totalizing servers.}
Let us measure how much it could affect the totals.

%
%
\begin{figure*}

\includegraphics{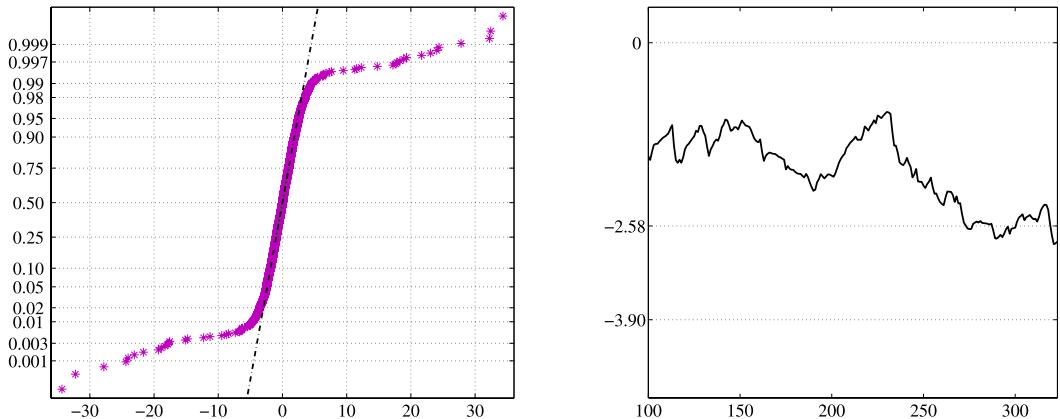}

\caption{Left panel: Normal probability plot of $Z^{\mathrm{OUT}}$ scores of
the presidential elections of 2000. Right panel: $\zeta_k$ versus $k$
for presidential elections of 2000.}\label{fig12}
\end{figure*}

Let $R$ be the population ratio defined in (\ref{pr}).
Bounds for the relative error, introduced in the vote counting by the
irregularities, 
can be obtained maximizing and minimizing the relative error $(r_k- R)/R$.
Thus, we can provide a prediction interval for the \textit{corrected
proportion} of votes in favor of Ch\'avez, namely,
\begin{eqnarray}\label{intervalestimator}
&&\biggl[\max_{100\leq k\leq K/2} \biggl(1-\frac{r_k -R}{R}\biggr) R
,\nonumber\\ [-8pt]\\ [-8pt]
&&\quad \min
_{100\leq k\leq K/2}\biggl(1-\frac{r_k -R}{R}\biggr) R \biggr],\nonumber
\end{eqnarray}
$K$ being the total number of notebooks. Note that we are considering
up to 50\% of the notebooks $(k\ll K)$,
those with the highest $|Z^{\mathrm{OUT}}|$ values.

For example, the prediction interval (\ref{intervalestimator}) for the
presidential election of 1998, in which Ch\'avez won with 56\% of valid
votes, is $[55\%, 57\%]$. We remark that this is an example of an
atypical but fair election, where the results were well accepted by
political parties and international observers.

Let us consider next the presidential elections of 2000.
As mentioned above, The Carter Center considers this election as flawed
and not fully successful.
However, they also emphasize that the irregularities did not change the
presidential results.
Our methodology confirms their conclusion.
Figure \ref{fig12} summarizes our testing analysis.
We observe the highest presence of $Z^{\mathrm{OUT}}$ outliers in 2000:
9\% of $Z^{\mathrm{OUT}}$ values (327 of 3730) are outside the 99\% normal
confidence interval.
Also, the most extreme $Z^{\mathrm{OUT}}$ scores of 2000 are higher than the
observed in 1998 and 2004.
But, there is not evidence to reject $\mathcal{H}_1$ for this election.
The $\zeta$ scores series is always in the 99\% normal confidence
interval, except for a short excursion.
Moreover, the prediction interval (\ref{intervalestimator}) for this
election is $[59\%, 62\%]$, and Ch\'avez was elected with 59\% of the
valid votes.

We do observe a controversial result in the referendum, managed by a
different electoral umpire from those that managed the elections of
1998 and 2000: The prediction interval is $[47\%, 57\%]$. The official
result (59\%) is out of range, while results that overturn the winner
are within. We remark that while this is not proof that irregularities
changed the overall results, it does illustrate that such a scenario is
plausible. Certainly, the result should be, at least, more in line with
the prediction interval.




\section{Conclusions}\label{sec4}

The main tool for conciliating political actors in an election under
suspicion of fraud is a full audit.
When this is not possible, statistical methods for detecting numerical
anomalies and diagnosing irregularities can be useful for evaluating
the likelihood of the allegations of fraud.
This is the aim of \textit{election forensics} (Mebane, \citeyear
{mebane2008}), an exciting area of applied statistics.
Election forensics has been applied for several recent controversial
elections, including 2004 USA, 2006 Mexico, 2008 Russia and 2009 Iran
(Mebane, \citeyear{mebane2009});
see the personal web page of Walter Mebane.\footnote{\href{http://www-personal.umich.edu/\textasciitilde wmebane}{www-personal.umich.edu/\textasciitilde wmebane}.}
The Venezuelan recall referendum is a case study that shows a wide
pallet of the commonly used statistical tools and problems that can
arise in this type of analysis,
as shown by our review in Section \ref{sec2}.
In particular, we have highlighted problems related to exit polls,
causal relationship between number of votes and dependent variables,
Benford's Law, different levels of data aggregation, goodness of fit,
and election modeling.
Beyond the statistical learning, the hard criticism of some of the
papers reviewed relates to a deep concern about the future~of this
emerging area.
I am convinced that the diffusion of~inaccurate analyses only causes
founded allegations of fraud to be undervalued.
At least, this was the case~of the Venezuelan referendum.

We propose a forensic election methodology, based only on vote counting,
to analyze the referendum.
Also the Venezuelan presidential elections of 1998 and 2000 are reviewed.
Unlike previous work, we used the full information of the official dataset.
This consists not only of the number of votes for and against revoking
the mandate of President Chavez, but also the number of abstentions and
invalid votes at the official data unit with the lowest number of votes.
The main conclusion of the present paper is that there were a
significant number of irregularities in the vote counting that
introduced a bias in favor of the winning option. We provide prediction
intervals for the bias, showing that the scenario in which the bias
could overturn the results is plausible. This places solid evidence in
the arena, substantiating the allegations of fraud made at the time.

\section*{Acknowledgments}
The author acknowledges the inspiring discussions
on the subject with Hayd\'ee Lugo, of Universidad Complutense
de Madrid. Mark Lindeman suggested the hypothesis discussed in
Section \ref{sec36}. The author thanks the associate editor and the
anonymous reviewers, for a very careful reading of the manuscript
and thoughtful comments.
Supported in part by\break Spanish MEC Grant ECO2011-25706
and CAM\break Grant S2007/HUM-0413.


\end{document}